\documentclass[aps,pre,reprint,nofootinbib,floatfix]{revtex4-1}
\usepackage{graphicx,color,psfrag}
\newcommand{\alb}[1]{#1}
\usepackage{amsfonts}
\usepackage{dcolumn}% Align table columns on decimal point
\newcommand{\E}{\mathrm{e}}
\newcommand{\average}[1]{\left<{#1}\right>}
\newcommand{\p}[1]{\left({#1}\right)}
\newcommand{\pq}[1]{\left[{#1}\right]}

\newcommand{\D}{\mathrm{d}}

\newcommand{\ii}{\mathrm{i}}
\newcommand{\Jss}{\bar J}
\newcommand{\bs}{\bf s}
\newcommand{\bh}{{\mathbf h}}
\begin{document}
\title{A minimal model of an autonomous thermal motor}
\author{ Hans C. Fogedby and Alberto Imparato}
\affiliation{Department of Physics and
Astronomy, University of
Aarhus, Ny Munkegade\\
8000 Aarhus C, Denmark\\}
\begin{abstract}
We consider a model of a Brownian motor composed of two coupled overdamped degrees of freedom
moving in periodic potentials and driven by two heat reservoirs. This model exhibits a spontaneous breaking of symmetry  and gives rise to directed transport in the case of a non-vanishing interparticle interaction strength. For strong coupling between the particles we derive an expression for the propagation velocity
valid for arbitrary periodic potentials. In the limit of strong coupling the model is equivalent to the 
B\"uttiker-Landauer model  \cite{Landauer1988,Hondou2000,Benjamin08} for a single particle diffusing in an environment with position dependent temperature. By using numerical calculations
of the Fokker-Planck equation and simulations of the Langevin equations we study the model for arbitrary
coupling, retrieving many features of the strong coupling limit. In particular, directed transport emerges even for symmetric potentials. For distinct heat reservoirs the heat currents are well-defined quantities allowing a study
of the motor efficiency. We show that the optimal working regime occurs for moderate coupling.
Finally, we introduce a model with discrete phase space which captures the essential features
of the continuous model, can be solved in the limit of
weak coupling, \alb{ and exhibits a larger efficiency than the continuous counterpart}.
\end{abstract}
.
%\date{\today}

\maketitle
%%%%%%%%%%%%%%%%%%%%%%%%%%%%%%%%%%%%%%%%%%%%%%%%%%%%%%%%%
%%%%%%%%%%%%%%%%%%%%%%%%%%%%%%%%%%%%%%%%%%%%%%%%%%%%%%%%%
{\it Introduction}
%%%%%%%%%%%%%%%%%%%%%%%%%%%%%%%%%%%%%%%%%%%%%%%%%%%%%%%%%
%%%%%%%%%%%%%%%%%%%%%%%%%%%%%%%%%%%%%%%%%%%%%%%%%%%%%%%%%

There is currently numerous scientific investigations aimed at characterizing the functioning of micro and nano-motors. There has, for example, been a rapid development of various artificial nanomotors
with the aim of mimicking the performance of biological machines \cite{Kay2007,Liu2009,Lund2010}.
 
From the point of view of man-made engineered micro and nano-motors, ideally one would like to design autonomous machines which are able to cyclically extract energy from the resources available in the environment 
and convert it to useful work. Similarly to their macroscopic counterparts, such machines must be driven out-of--equilibrium by means of one or more thermodynamic forces. 

In the present paper we focus in particular on a motor driven by temperature gradients.
A Brownian motor has long been the paradigmatic model for a microscopic machine, working either in time-dependent or  steady state conditions. One well known example is a  Brownian particle moving in a periodic and asymmetric potential, a so-called ratchet potential. In such a spatially periodic system, the breaking of the spatial inversion symmetry and of  thermal equilibrium, obtained by modulating the force acting on the particle, 
 results in the emergence of directed transport \cite{Reimann02,VandenBroeck08,Gehlen09}.
Another typical example is represented by a Brownian particle driven by both a periodic temperature variation and an external parameter, periodically changing the system energy \cite{Schmiedl08,Brandner15}. This model,
which mimics the operation of a heat engine  cyclically in contact with different heat reservoirs, has been implemented in a recent experiment \cite{Blickle2012}. In all these models there is an external agent that changes periodically some parameters, typically a thermodynamic force, according to the motor state in its phase space.

However, the optimal design for a thermal engine is achieved by an autonomous motor which can operate in steady state conditions without any external time dependent drive.
A well known example of an autonomous motor is the so called B\"uttiker-Landauer
model  \cite{Landauer1988,Hondou2000,Benjamin08}, consisting of  a Brownian particle moving in a periodic 
potential and a periodic temperature profile. In this model the spatial symmetry is broken by a phase shift between the potential and the temperature profile  \cite{Matsuo2000}, resulting in a direct particle current. However, for such a system the definition of efficiency presents an issue \cite{Berger09}, e.g., the heat transfer cannot be evaluated without ambiguity in the overdamped regime \cite{Hondou2000}.
Still, the most remarkable example of autonomous design is the Feynman ratchet \cite{Feynman:1963}, where 
both spatial symmetry and thermal equilibrium are explicitly broken. In the context of Brownian motion, such a ratchet has been modelled, for example, with asymmetric objects moving in separate thermal baths \cite{VandenBroeck04,Fruleux12,Holubec17}.
\alb{Another class of autonomous machines is represented by Brownian gyrators \cite{Reimann07,Dotsenko13,Argun17} where a Brownian particle in a parabolic asymmetric potential rotates around potential minimum  when connected to two different heat reservoirs. The particle mean rotation velocity in the phase space can be calculated \cite{Dotsenko13},  although the problem of how to extract useful work from such a setup has not been addressed.}
 In \cite{Sancho05} the authors introduced a Brownian motor consisting of two Brownian particles  with linear and strong coupling maintained at different temperatures and moving in asymmetric ratchet potentials, so as to mimic the asymmetric features of the classical Feynman ratchet and pawl system.

In the present paper, inspired by the last model above, we present a minimal model of  an autonomous thermal motor composed of two Brownian particles moving in two (possibly symmetric) periodic potentials, interacting with a general periodic potential,  and maintained at different temperatures.
We show that such a system does not require ratchet potentials (with, e.g., an  asymmetric saw-tooth shape) in order to exhibit directed transport, the spatial symmetry being broken by the interaction between the particles.
We solve the model analytically in the strong coupling limit for general potentials and show that in this limit the model is equivalent to  
the B\"uttiker-Landauer
model  \cite{Landauer1988,Hondou2000, Reimann02, Benjamin08}.
We study the model by numerically solving the Fokker-Planck equation and by numerical integration of the Langevin equation for arbitrary coupling strength, and investigate the dependence of the system velocity on the relevant set of parameters. We show that the particle current arises as soon as there is a non-vanishing coupling between the particles, and find that several features of the strong coupling limit are also present  in the weak to moderate coupling regime.
We derive an expression for the heat current and, by applying an external force,  also evaluate the motor thermodynamic efficiency. 
Our results indicate that the optimal regime, as far as the motor velocity and efficiency are concerned, occurs in the moderate coupling regime.
We finally introduce and discuss a minimal discrete model, that can be solved exactly for any coupling, in particular we obtain the exact expression for the motor current and the heat currents. Such expressions corroborate our findings for the continuous model in the weak coupling regime.
%Our work has been inspired by Gomez-Marin and J. M. Sancho \cite{Sancho05}, which considered a  model of Brownian motor with two particles in periodic and asymmetric potentials with given shapes, with a linear interaction and  in the limit of strong coupling.

%%%%%%%%%%%%%%%%%%%%%%%%%%%%%%%%%%%%%%%%%%%%%%%%%%%%%%%%%
%%%%%%%%%%%%%%%%%%%%%%%%%%%%%%%%%%%%%%%%%%%%%%%%%%%%%%%%%
{\it Model}
%%%%%%%%%%%%%%%%%%%%%%%%%%%%%%%%%%%%%%%%%%%%%%%%%%%%%%%%%
%%%%%%%%%%%%%%%%%%%%%%%%%%%%%%%%%%%%%%%%%%%%%%%%%%%%%%%%%

The model consists of two overdamped coupled degrees of freedom moving in periodic potentials and driven by two
heat reservoirs maintained at different  temperatures $T_1$ and $T_2$.  Denoting the degrees of
freedom by $x_1$ and $x_2$, the  model is characterized by the potential
\begin{eqnarray}
V(x_1,x_2)=V_1(x_1)+V_2(x_2)+k u(x_1-x_2),
\label{pot}
\end{eqnarray}
where $V_i$ are periodic potentials with period $L_i$, $i=1,2$,
and $u(x_1-x_2)$ a periodic interaction
potential, with interaction strength $k$ and period $L_u$.
We assume that the periods $L_i$ and $L_u$ are commensurable, such that $L=\max(L_1,L_2,L_u)$ is the total potential period, and $L=n L_1=m L_2=l L_u$, with $n,\, m,\, l$ integer numbers.
 Setting the friction constant $\Gamma=1$ and denoting the forces
by $F_i=-dV_i/dx_i$ the overdamped coupled Langevin equations have the
form (a dot denoting a time derivative, a prime denoting a space derivative)
\begin{eqnarray}
&&\dot x_1=F_1(x_1)-k u'(x_1-x_2)+\eta_1(t),
\label{lang1}
\\
&&\dot x_2=F_2(x_2)-ku'(x_2-x_1)+\eta_2(t);
\label{lang2}
\end{eqnarray}
here the white Gaussian noises $\eta_1$ and $\eta_2$, characterizing the heat reservoirs 
at temperatures $T_1$ and $T_2$, are correlated according to $\langle\eta_i(t)\eta_j(t')\rangle=2T_i \delta_{ij}\delta(t-t')$.
%
%\begin{eqnarray}
%&&\langle\eta_1(t)\eta_1(t')\rangle=2T_1\delta(t-t'),
%\label{noise1}
%\\
%&&\langle\eta_2(t)\eta_2(t')\rangle=2T_2\delta(t-t').
%\label{noise2}
%\end{eqnarray}
%
 %For $k=0$ the particle are uncoupled and equilibrate individually with joint distribution
 %$P(x_1,x_2)\propto\exp(-V_1(x_1)/T_1)\exp(-V_2(x_2)/T_2)$. For $T_1=T_2=T$ the system is in equilibrium
 %with distribution $P(x_1,x_2)\propto\exp(-U(x_1,x_2)/T)$. 
In the non equilibrium case for $T_1\neq T_2$
 a heat flux is established between the reservoirs. We show that if the following conditions are met $i)\, k\neq0$ and $ii)\, V_1\neq V_2$ , the system behaves as a motor and part of the integrated heat flux is used to sustain a non-vanishing velocity of the center of mass $\bar v$.  In the following we will give a precise formulation of the condition  $V_1\neq V_2$. 

According to the standard definition in stochastic thermodynamics \cite{Sekimoto:Book}, the rate of heat exchanged with each reservoir along a single stochastic trajectory is $\dot Q_i=\dot x_i(t)\partial_iV(x_1,x_2)$. Using a standard approach \cite{Imparato07,Fogedby12,Fogedby14} we then obtain the average heat rate
\begin{equation}
\average{\dot Q_i}=\average{T_i \partial_i^2V(x_1,x_2)-\p{\partial_i V(x_1,x_2)}^2};
\label{dotQ}
\end{equation} 
see appendix \ref{app2} 
for the details of the calculation.

In order to evaluate the thermodynamic efficiency of the motor, we apply a force $f_i$ to one of the particles and choose the sign of $f_i$ such that the force opposes the center of mass motion, whose direction we assume as the positive one. The Brownian motor will thus do work against the external force and the corresponding output power is $-f_i  v_i$.
Consequently, the efficiency is given by
\begin{equation}
\eta=-f_i v_i/\average{\dot Q_H},
\label{eta:eq}
\end{equation} 
where the index $H$ labels the hot reservoir.

 %%%%%%%%%%%%%%%%%%%%%%%%%%%%%%%%%%%%%%%%%%%%%%%%%%%%%%%%%
%%%%%%%%%%%%%%%%%%%%%%%%%%%%%%%%%%%%%%%%%%%%%%%%%%%%%%%%%
{\it Analysis for large $k$}
%%%%%%%%%%%%%%%%%%%%%%%%%%%%%%%%%%%%%%%%%%%%%%%%%%%%%%%%%
%%%%%%%%%%%%%%%%%%%%%%%%%%%%%%%%%%%%%%%%%%%%%%%%%%%%%%%%%

 The coupled Langevin equations (\ref{lang1}) and (\ref{lang2}) as well as the associated Fokker-Planck equation
 are difficult to analyze. However, in the adiabatic strong coupling limit for large $k$ the model is amenable
 to analysis; details of the calculations are reported in appendix \ref{app1}. 
Following  \cite{Sancho05} we 
 note  that the relative coordinate $y=(x_1-x_2)/2$ is suppressed and
 its dynamics quenched, i.e., $y\sim 0$ and $\dot y\sim 0$. Moreover, introducing also the center of mass 
 coordinate $x=(x_1+x_2)/2$, setting $\dot y=0$, and eliminating the fast variable $y$,  
 we obtain a single Langevin equation for $x$,
\begin{equation}
\dot x=h(x)+g(x)\xi(t),
\label{lan}
\end{equation} 
with $\xi(t)$ a Gaussian white noise, $\langle\xi(t)\xi(t')\rangle=2\delta(t-t')$.
Here the drift term $h$ is given by 
\begin{equation}
h(x)=F_1(x)s_1(x)+F_2(x)s_2(x),
\label{h}
\end{equation} 
where the space dependent diffusion coefficient $g^2$ depends on the reservoir
temperatures and on the particle potentials. It has the form
\begin{eqnarray}
g^2(x)&=&T_1s_1(x)^2+T_2s_2(x)^2,
\label{g}\\
s_{1,2}(x)&=&\frac{2k-F_{2,1}'(x)}{4k-(F_1'(x)+F_2'(x))}.
\label{si}
\end{eqnarray}
From the definitions it follows that the drift and diffusion are periodic functions of $x$ with period $L$.
For a constant $g=\sqrt T$, $T=(T_1+T_2)/2$, the Langevin equation (\ref{lan}) describes a Brownian particle 
subject to the force $h(x)$.  However, for a periodic "temperature"
$T(x)=g(x)^2$ the Langevin equation  exhibits the "blow torch" effect as in the 
 B\"uttiker-Landauer model \cite{Landauer1988,Benjamin08} and thus  give rise to a motor effect, as detailed below.

In order to determine the center of mass velocity $\bar v=\langle\dot x\rangle$ 
we consider the non linear Langevin equation (\ref{lan}) driven by multiplicative noise
$g(x)\xi(t)$ and derive the associated Fokker-Planck (FP) equation
\cite{Risken}.
Adhering to the Stratonovich interpretation the FP equation has the form $dP/dt=-dJ/dx$, where the 
probability current is given by $J(x)=(h(x)-g(x)g(x)')P(x)-g^2(x)P'(x)$.
%
%\begin{eqnarray}
%J(x)=(h(x)-g(x)g(x)')P(x)-g^2(x)P'(x).
%\label{cur}
%\end{eqnarray}
%
%The issue is thus to determine a stationary distribution $P(x)$ which is periodic with period $L$, corresponding to a constant current $J$, and thus to a non zero propagation velocity $\bar v=L J $. 
The $L$-periodic stationary solution of the FP equation reads
\begin{equation}
P(x)=\bar J \frac{\E^{-U(x)}}{(1-\E^{\bar f L}) g(x)}\int_x^{x+L}  \frac{\E^{U(y)}}{g(y)}\,  d y,
\label{eqP:ss}
\end{equation} 
where we have introduced the effective potential 
\begin{equation}
U(x)=-\int_0^x d y\, h(y)/g^2(y).
\label{eq:U}
\end{equation} 
The normalization condition $\int_0^L dxP(x)=1$ then yields the constant steady-state current
\begin{eqnarray}
\bar J=(1-\E^{\bar f L})\pq{\int_0^Ldx \frac{\E^{-U(x)}}{g(x)}\int_x^{x+L}dy \frac{\E^{U(y)}}{g(y)}}^{-1},
\label{cur2}
\end{eqnarray}
and thus the non zero propagation velocity $\bar v=L \bar J$. Here the quantity $\bar f=[U(x+L)-U(x)]/L$ quantifies the breaking of the right-left symmetry.
The expression (\ref{cur2}) for the current is a central result.
We infer that although  $h(x)$ and $g(x)$ are periodic function the average $h(x)/g^2(x)$ over one period, as given by eq.~(\ref{eq:U}), must be  non-vanishing in order to ensure directed transport. The condition that $\bar f \neq 0$ in order for the present model to exhibit direct transport is the same as in the   B\"uttiker-Landauer model for a single particle in a force field $h(x)$ and a position dependent profile $T(x)=g^2(x)$ \cite{Matsuo2000}.

Expressing the potentials in their Fourier representation 
%$V_1(x)=\sum_n c_n\exp(2\pi inx/L)$ and $V_2(x)=\sum_nd_n\exp(2\pi inx/L)$
$V_i(x)=\sum_{q_i} v_{i,q_i} \exp(\ii q_i x)$, with $v^*_{i,q_i}=v_{i,-q_i}$ and $q_i=2 \pi n_i/L_i$,
and evaluating the ratio $ h(y)/g^2(y)$ in equation (\ref{eq:U}) to leading order in $1/k$, we obtain 
\begin{equation}
U(x)=U_0(x)- x \bar f,
\label{eq:Utilt}
\end{equation} 
where $U_0(x)$ is a $L$-periodic potential, that can be written in terms of the Fourier components of the two potentials $V_i(x)$ and of the two temperatures $T_{1,2}$, while for $\bar f $ we obtain
\begin{eqnarray}
\bar f = -2  \frac{(T_1-T_2)}{k (T_1+T_2)^2} \sum_q q^3\mathrm{Im}\p{v_{1,q} v^*_{2,q}};
\label{del22}
\end{eqnarray}
see appendix \ref{app1} 
for the details.
Inspection of eq.~(\ref{eq:Utilt}) suggests that the quantity $\bar f$ plays the role of a constant tilting force for the periodic potential $U_0(x)$, as found in models of isothermal molecular motors \cite{Golubeva2012,VandenBroeck2012}, where a  Brownian particle moves in a tilted periodic potential.
By inspection of eq.~(\ref{del22}) we observe that for general unequal periodic potentials the  necessary conditions for 
$\bar f\neq 0$ are $a)$ $T_1\neq T_2$ and $b)$ at least one common mode of the two potentials. Furthermore, if the potentials $V_1$ and $V_2$ are identical but  shifted with respect to one another, $V_2(x)=V_1(x+\phi)$, we find 
\begin{eqnarray}
\bar f= 2  \frac{(T_1-T_2)}{k (T_1+T_2)^2} \sum_q q^3 |v_{1,q}|^2\sin(q \phi),
\end{eqnarray}
implying that the current and thus the steady state velocity in this case is non-zero if, for at least one mode in the potential decomposition, $\phi q\neq \pi m$, with $m$ integer.

{\it Arbitrary coupling strength}

\begin{figure}[h]
\center
\psfrag{d=1}[lt][lt][.8]{$n=1$}
\psfrag{d=2}[lt][lt][.8]{$n=2$}
\psfrag{d=4}[lt][lt][.8]{$n=4$}
\psfrag{k}[ct][ct][1.]{$k$}
\psfrag{Q2}[ct][ct][1.]{$\average{\dot Q_2}$}
\psfrag{v}[ct][ct][1.]{$\bar v $}
\includegraphics[width=8cm]{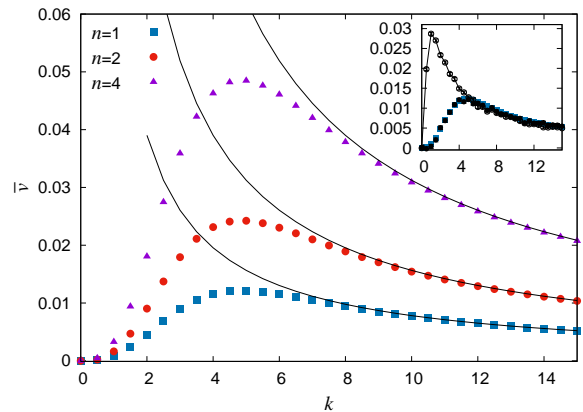}
\caption{Velocity $\bar v$, as defined by eq.~(\ref{v:smo}), as a function of the interaction strength $k$  for the potential (\ref{pot1}), with $u(z)= -\cos(n_u z)$, $n=n_1=n_2=n_u$, $T_1=1, \, T_2=2.5,\, \varphi =\pi/2$, $a_1=a_2=1$. The full lines correspond to the analytic solution  $\bar v=L \bar J$ in the limit of large $k$ with $\bar J$ given by eq.~(\ref{cur2}). Inset: Comparison with numerical simulations for $n=1$. The error bar points are obtained by numerical integration of the Langevin eqs.~(\ref{lang1})-(\ref{lang2}), with $10^4$ independent trajectories. \alb{Filled circles: periodic interaction potential $u(z)=-\cos(z)$. Open circles: quadratic interaction potential $u(z)=z^2/2$, the line is a guide to the eye . In the limit of large $k$ the two potentials give the same velocity, since the relative coordinate $x_1-x_2$ is small, and one can make a quadratic approximation for the periodic potential $-\cos(x_1-x_2)\simeq (-1+(x_1-x_2)^2/2)$ }.}
\label{fig:num}
\end{figure}

In the case of arbitrary coupling strength $k$ and general periodic potentials in eq.~(\ref{pot}) a numerical 
solution of the Fokker Plank equation in the long time limit yields the steady state PDF $P_{ss}(x_1,x_2)$.
%In principle, the same procedure can be applied to any potential form in eq.~(\ref{pot}).
The steady state velocity is then obtained from eqs.(\ref{lang1})-(\ref{lang2}) according to
\begin{equation}
\bar v=\frac 1 2 \average{\dot x_1+\dot x_2}=\frac{1}{2} \average{F_1(x_1)+F_2(x_2)},
\label{v:smo}
\end{equation}  
where the last average is calculated with respect to  $P_{ss}(x_1,x_2)$. We have, moreover,  corroborated our findings 
by means of direct numerical simulations of the Langevin equations~(\ref{lang1})--(\ref{lang2}).
In the following we choose the potential 
\begin{equation}
V(x_1,x_2)=a_1 \cos(n_1 x_1)+a_2\cos(n_2 x_2+\varphi)+k u(x_1-x_2), 
\label{pot1}
\end{equation} 
with $u(z)=-\cos(n_u z)$ if not otherwise stated,  and with arbitrary coupling strength $k$. 
We notice that while each single contribution on the rhs of eq.~(\ref{pot1}) is a symmetric function, the total potential is not.
We commence our analysis by considering the case where the three terms in the potential  (\ref{pot1}) have the same period. The results are shown in fig.~\ref{fig:num}.

\begin{figure}[h]
\center
\psfrag{k=2.5}[ct][ct][.9]{$k=2.5$}
\psfrag{k=5}[ct][ct][.9]{$k=5\quad$ }
\psfrag{k=10}[ct][ct][.9]{$k=10\,\, $}
\psfrag{phi}[ct][ct][1.]{$\varphi$}
\psfrag{v}[ct][ct][1.]{$\bar v $}
\psfrag{T2}[ct][ct][1.]{$T_2$}
\includegraphics[width=8cm]{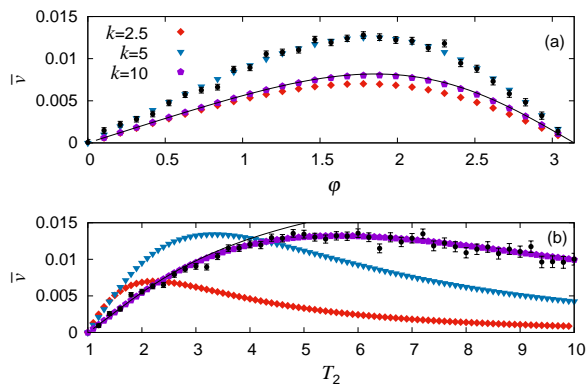}
\caption{Panel a): steady state velocity $\bar v$ (\ref{v:smo}) as a function of the phase shift  $\varphi$  for the potential (\ref{pot1}) with $n_1=n_2=n_u=1$, $T_1=1, \, T_2=2.5$, $a_1=a_2=1$, and different values of the coupling constant $k$. The error bar points are obtained by numerical integration of the Langevin eqs.~(\ref{lang1})-(\ref{lang2}), with $10^4$ independent trajectories.   The full line corresponds to the analytic solution  $\bar v=L \bar J$ in the limit of large $k$ with $\bar J$ given by eq.~(\ref{cur2}). Panel b): Steady state velocity $\bar v$  as a function of the temperature $T_2$,  for the potential (\ref{pot1}) with $n_1=n_2=n_u=1$, $T_1=1,\, \varphi=\pi/2$, and different values of the coupling constant $k$. Symbols as in panel (a). }
\label{fig1:num}
\end{figure}

We find excellent agreement with the large $k$ result discussed above, while for fixed $k$ the velocity increases with the potentials common frequency.
As anticipated, the optimal velocity is obtained in the moderate coupling strength regime.
Next we consider the cases where the coupling $k$ is fixed and we change $a)$ the phase $\varphi$ between the two potentials $V_i(x_i)$ and $b)$ the temperature difference, see fig.~\ref{fig1:num}. 

As in the case of a large coupling strength, we find that if the two potentials are identical with no phase shift, the center-of-mass velocity vanishes.

As anticipated  the velocity vanishes for $T_1=T_2$, independently of $k$.
The optimal temperature bias $T_1-T_2$ depends on the coupling strength $k$ and the largest value of the velocity is achieved in the moderate coupling regime. For large values of the temperature difference the thermal fluctuations become too large to favour a coordinated  motion of the center of mass in a given direction.

\begin{figure}[h]
\center
%\psfrag{k}[ct][ct][1.]{$k$}
%\psfrag{tau1}[ct][ct][1.]{$f_1$}
\includegraphics[width=8cm]{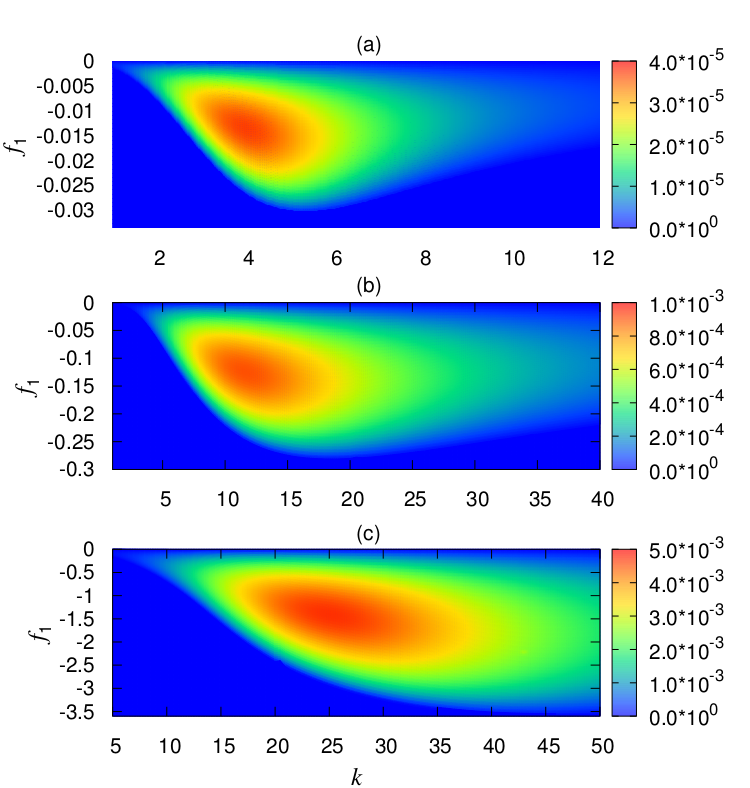}
\caption{Efficiency $\eta$ in eq.~(\ref{eta:eq}) as a function of the interaction strength $k$ and of the external force $f_1$,  for the potential (\ref{pot1}), with different frequencies between the interaction and the particles' potential.  $T_1=1, \, T_2=2.5$ and $n_u=1$. (a) $a_1=a_2=1$, $n_1=n_2=1$. (b) $a_1=a_2=1$, $n_1=n_2=4$. (c) $a_1=a_2=5$, $n_1=n_2=4$.}
\label{fig:eta_3D}
\end{figure}

Applying a force $f_1$ to particle 1 we evaluate the efficiency using eq.~(\ref{eta:eq}). The results for $n_1=n_2=n_u=1$ are shown in fig.~\ref{fig:eta_3D}-(a). We observe that the maximal efficiency one can achieve with this set of parameters is quite small, of the order of $4\times 10^{-3}\, \%$. As long as the three potentials in eq.~(\ref{pot}) have  the same period $L$,
 changing $L$ corresponds to rescaling the single unit length, and thus the velocity $\bar v$ will decrease linearly with the potential period, while the heat rate (\ref{dotQ}) scales as $1/L^2$, see appendix \ref{app3}.
Thus, one cannot improve the motor  maximal efficiency at constant $k$ just by changing the common period $L$.
Inspection of equation (\ref{del22}) suggests that, in the strong coupling limit, the contribution of each harmonic to  the linear tilt in the effective potential $U (x)$
scales as $q^3$ at constant $k$. This suggests a strategy to  enhance the velocity and thus possibly the efficiency.
In the following we will thus evaluate the efficiency $\eta$ by fixing the period of the interacting potential $u(z)$ and increase the period of the two potentials $V_1(x_1)$ and $V_2(x_2)$.
The results for a given choice of parameters  are shown in fig.~\ref{fig:eta_3D}, and we find indeed an increase in $\eta$ with a maximal value of the order of 0.1 \%, any further increase in $n_1=n_2$ does not give rise to a higher maximal value of $\eta$ (data not shown).

\alb{ Another possible strategy to increase the system efficiency is to increase the amplitudes $a_1,\, a_2$  of the potentials $V_1(x_1)$ and $V_2(x_2)$  so as to allow the motor to sustain a larger force before reaching the stall condition, while keeping the fluctuations of the relative coordinate small, so as to reduce the heat currents, and thus the denominator in eq.~(\ref{eta:eq}). While this approach does increase the efficiency, see figure \ref{fig:eta_3D}-c, the achieved values are still quite small. }

\alb{In general our results show that the efficiency of the model motor is not very high. This is due to the lack of strong coupling between the input and the output energy current: the heat currents flow between the two reservoirs even when the center-of-mass wanders about a given position, without advancing in the positive direction. We discuss this point in details below, when comparing the efficiency of the continuous and the discrete models.}

\alb{We want to stress that the requirement for the interaction potential  $u(z)$ to be periodic is not necessary for the system to exhibit a non-zero velocity. A quadratic potential $k u(z)=k z^2/2$ also results in directed motion, see inset in figure \ref{fig:num}, since the total potential  still breaks the left right symmetry. This is the case, for example, in the strong coupling regime, where the relative coordinate is small, and the cosine interaction potential can be approximated by a quadratic polynomial. 
%However by choosing the total potential $V(x_1,x_2)$ to be periodic, one obtains a compact expression for the heat rates eq.~(\ref{eta:eq}). Furthermore...
However by choosing a periodic total potential one can solve the FP equation for $P_{ss} (x_1 , x_2 )$ by imposing periodic boundary conditions. On the contrary, when taking a non-periodic $u(z)$ one is only left with the results of the numerical integration of the Langevin equations.}

\alb{It is interesting to draw an analogy between our periodic model in its simplest form eq.~(\ref{pot1}) and the Hamiltonian for the $xy$-model, describing the elastic free energy in ferromagnetic or liquid crystal systems \cite{Chaikin}. The $xy$-model Hamiltonian for a spin model on a plane %invariant under continuous rotation 
reads   $H=-J\sum_{<i,j>} \bs_i \cdot \bs_j-\sum_i \bh_i \cdot \bs_i$, with $\bs_i=(\cos \theta_i, \sin \theta_i)$. We now isolate the contribution from the spins $i=1$ and $j=2$, and take $\bh_1=(h_1,0)$ and  $\bh_2=(0,h_2)$.  We obtain 
 $H_{1,2}=-J \cos(\theta_1-\theta_2)-h_1 \cos(\theta_1)-h_2 \sin(\theta_2)$, which is equivalent to eq.~(\ref{pot1}), provided that one takes $h_1=h_2=-1$, $n_1=n_2=n_u=1$,   $\varphi=\pi/2$, and shifts the coordinate $x_2\to x_2-\pi$.
}

{\it A discrete model}

Here we introduce a model with a discrete phase space which captures the essential features of the model described above.
We consider two particles $n=1,2$ that can occupy  different positions $a\cdot i_n$ on two regular, periodic lattices, where $a$ is the common lattice step and $i_n=\dots -2,-1,0,1,2,\dots$
Each particle is in contact with a thermal reservoir at inverse temperature $\beta_n$. The particles thus have right and left transition rates $\omega^+_n$ and $\omega^-_n$, respectively. We also assume that the diffusion is unbiased when the particles are uncoupled, i.e., for $\omega^+_n=\omega^-_n=\omega_n^0$, corresponding to a vanishing average velocity.

Next we introduce an interaction periodic potential depending on the distance of the particles on the lattice. Without loss of generality we assume that $U$ is a $2 \pi-$periodic function of the particle distance $(i_1-i_2)a$, i.e.,
\begin{equation}
U(i_1,i_2)=\frac k 2 \cos\pq{(i_1-i_2)a+\varphi}.
\label{U:def}
\end{equation} 
By taking $a=2 \pi/N_s$, this corresponds to the energy of the clock model ( discrete xy-model)\cite{Chaikin} with  lattice spin variables constrained to point in one of the $N_s$ directions.
%If we choose the lattice step to be $a=\pi$, each particle can take on two different states  $i_n=0,1$ within one period. However, with this choice the interaction potential does not break the left right symmetry: for example the states $(0,0)$ has the same energy difference with the states  $(0,\pm1)$, where particle $n=2$ has moved a step to the right or to the left, regardless of the value of phase $\varphi$.
%Thus, our next choice is $a=2 \pi/3$, corresponding to  three states  $i_n=0,1,2$ within one period.
If we choose the lattice step to be $a=\pi$, the interaction energy can assume two different energy values, $U=\pm k\cos(\varphi)/2$.
However, with this choice the interaction potential does not break the left right symmetry:  $U(-i_1,-i_2)=U(i_1,i_2)$ for any value of the  phase $\varphi$.
Consequently,  our next choice is $a=2 \pi/3$, corresponding to  three degenerate values of the energy $U=k/2 \cos \varphi,\,  k/2 \cos( \varphi\pm 2 \pi/3)$ as long as $\varphi\neq0, \pi$. With this choice the system potential breaks the left right symmetry:   $U(-i_1,-i_2)\neq U(i_1,i_2)$ for $\varphi\neq 0,\, \pi$ \footnote{ More precisely, there is no translation distance $\Delta$ such that $U(-i_1,-i_2)= U(i_1+\Delta,i_2+\Delta)$,  but since
 the potential depends on the difference $i_1-i_2$  the condition reduces to the one above}.

To simplify the model we assume that only one particle at a time can jump left or right. We are thus left with the following 
choice of transition rates
$W(i_1,i_2\rightarrow i_1',i_2)$ and $ W(i_1,i_2\rightarrow i_1,i_2')$. These rates must be chosen such that when the two temperatures are equal $\beta_1=\beta_2=\beta$, the system reaches the thermal equilibrium state $P_{eq}(i_i,i_2)\propto \exp[\beta U(i_1,i_2)]$, with $i_n=0,1,2$.
We thus impose that the transition rates of each particle obey a local detailed balance condition dictated by the particle's reservoir of the form 
\begin{eqnarray}
\frac{W(i_1,i_2\rightarrow i_1',i_2)}{W(i'_1,i_2\rightarrow i_1,i_2)}&=&\E^{-\beta_1 \pq{U(i_1',i_2)-U(i_1,i_2)}},\label{db1}\\
\frac{W(i_1,i_2\rightarrow i_1,i_2')}{W(i_1,i'_2\rightarrow i_1,i_2)}&=&\E^{-\beta_2 \pq{U(i_1,i'_2)-U(i_1,i_2)}}. \label{db2}
\end{eqnarray} 

There are several choices that enforce this condition. In order to make the rates symmetric we choose 
\begin{eqnarray}
W(i_1,i_2\rightarrow i_1',i_2)&=&\omega_0\E^{-\beta_1 \pq{U(i_1',i_2)-U(i_1,i_2)}/2},\label{W1:def}\\
W(i_1,i_2\rightarrow i_1,i_2')&=&\omega_0\E^{-\beta_2 \pq{U(i_1,i'_2)-U(i_1,i_2)}/2}, \label{W2:def}
\end{eqnarray} 
where we have assumed identical microscopic transition rates, i.e.,  $\omega_0=\omega^0_1=\omega^0_2$; in the following we also take $\varphi=\pi/6$.
The system consists of 9 different states, and the steady state $P_{ss}(i_1,i_2)$ can be solved for, together with the currents.
We find that both  currents read
\begin{equation}
J_n=-\omega_0\frac{ ( \gamma_1-1) (\gamma_1 - \gamma_2) ( \gamma_2-1)}{3 (1 + \gamma_1 \gamma_2 + \gamma_1^2 \gamma_2^2)},
\label{curr:dis}
\end{equation} 
where 
\begin{equation}
\gamma_n=\exp( \beta_n \bar k); \quad \bar k=\cos(\pi/6)k/4.
\end{equation} 

Expanding in Taylor series for small $k$, one obtains
\begin{eqnarray}
%J_n=\frac { \omega_0 (T_1- T_2)}{9 T_1^2T_2^2} \pq{\bar k^3- \frac{ T_1^2+3 T_1 T_2+T_2^2}{4T_1^2T_2^2} \bar k^5} + O(k^7); \qquad \bar k=\cos(\pi/6)k/4.
J_n&=&-\frac { \omega_0 }{9 }\beta_1\beta_2 (\beta_1-\beta_2) \pq{\bar k^3- \frac1 4 (\beta_1^2+\beta_1\beta_2+\beta_2^2) \bar k^5} \nonumber\\
&&+ O(k^7).
\label{curr:dis:tay}
\end{eqnarray} 
The particle currents are  odd functions of $k$ because of the particular choice of the potential (\ref{U:def}); changing $k\rightarrow-k$ correspond to a shift of $\pi$ in one of the two coordinates $ i_n a$.
Changing the phase $\varphi$ changes the prefactors in the definition of $\bar k$ but not the current scaling behaviour.

In the limit of large $k$, eq.~(\ref{curr:dis}) becomes
\begin{equation}
J_n\simeq \frac{\omega_0} {3} \pq{\frac 1 {\gamma_1} -\frac 1 {\gamma_2}}.
\end{equation} 

One might have chosen another expression for the transition rates, e.g. the Glauber's rates \cite{Glauber63}. However, with this choice one does not obtain a compact expression for the current as in eq.(\ref{curr:dis}), but the scaling behaviour for small $k$ is the same as in eq.~(\ref{curr:dis:tay}) with the leading order being proportional to $k^3$.
This is compatible with  what we find numerically for the continuous model, where the velocity $\bar v $ appears to have a vanishing first derivative for $k=0$, see fig.~\ref{fig:num}.

We can now calculate the entropy flow rate for each reservoir \cite{Lebowitz1999,Imparato7a,Seifert2012}
\begin{eqnarray}
\dot S_{p,1}=\sum_{i_1,i'_1=0,1,2} && W(i_1,i_2\rightarrow i_1',i_2) P_{ss}(i_1,i_2) \nonumber \\
&& \times \ln\frac{ W(i'_1,i_2\rightarrow i_1,i_2)} { W(i_1,i_2\rightarrow i_1',i_2)},\nonumber\\
\dot S_{p,2}=\sum_{i_2,i'_2=0,1,2} &&W(i_1,i_2\rightarrow i_1,i'_2) P_{ss}(i_1,i_2) \nonumber \\
&&  \times  \ln\frac{ W(i_1,i'_2\rightarrow i_1,i_2)} { W(i_1,i_2\rightarrow i_1,i'_2)},\nonumber
\end{eqnarray} 
and given that the transition rates (\ref{W1:def})-(\ref{W2:def}) obey the detailed balance condition for each reservoir, we obtain 
\begin{eqnarray}
T_1 \dot S_{p,1}= \average{\dot Q_1}= &&  \hspace{-.7cm}\sum_{i_1,i'_1=0,1,2}W(i_1,i_2\rightarrow i_1',i_2) P_{ss}(i_1,i_2)\nonumber \\
&& \qquad\times   \pq{U(i'_1,i_2) - U(i_1,i_2)},\label{q:disc1}\\
T_2 \dot S_{p,2}=\average{\dot Q_2}=&&\hspace{-.7cm}\sum_{i_2,i'_2=0,1,2} W(i_1,i_2\rightarrow i_1,i'_2) P_{ss}(i_1,i_2)\nonumber \\
&& \qquad\times  \pq{U(i_1,i'_2) - U(i_1,i_2)},\label{q:disc2}
\end{eqnarray} 
yielding
\begin{eqnarray}
\average{\dot Q_1}&=& 3\omega_0 k \frac{  (\gamma_1-\gamma_2)(1 +2 (\gamma_1+\gamma_2) + \gamma_1\gamma_2)}{1+  \gamma_1\gamma_2+ ( \gamma_1\gamma_2)^2},
%\average{\dot Q_2}&=& -3\omega_0 k \frac{  (\gamma_1-\gamma_2)(1 +2 (\gamma_1+\gamma_2) + \gamma_1\gamma_2)}{1+  \gamma_1\gamma_2+ ( \gamma_1\gamma_2)^2}.
\end{eqnarray} 
with $\average{\dot Q_2}=-\average{\dot Q_1}$.

Expanding to leading order in $k$ we find
\begin{equation}
\average{\dot Q_1}=4 \omega_0k^2 (\beta_1-\beta_2)+   O(k^4).
\end{equation} 

%{\it if I take the transition rates with an external force, the expression for the currents becomes messy and we don't learn much, I can make plots, but for that we already have the continuous model.}

\begin{figure}[h]
\center
%\psfrag{k}[ct][ct][1.]{$k$}
\includegraphics[width=8cm]{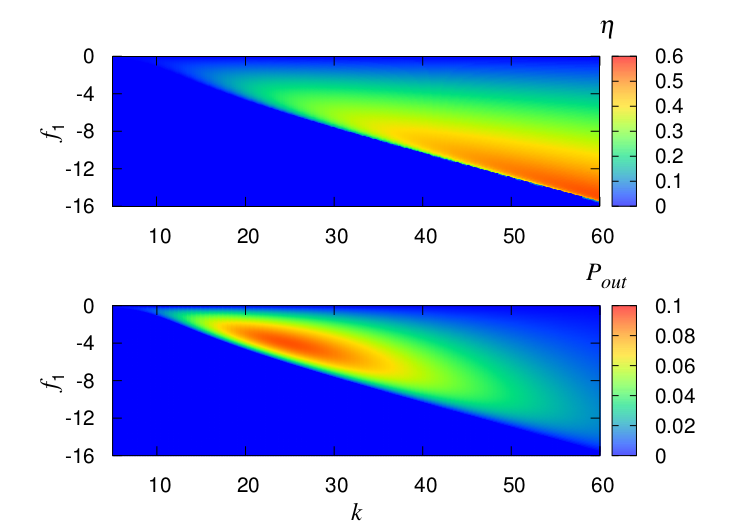}
\caption{\alb{Efficiency (top) and output power (bottom) for the discrete model as functions of the coupling strength $k$ and of the applied force $f_1$,  with $T_1=2.5$, $T_2=1$, $N_s=3$,  $\varphi=\pi/6$. The time scale is set such that $\omega_0=1$.}  }
\label{fig:disc}
\end{figure}

\alb{One can evaluate the efficiency by following the same approach as for the continuous model. By applying a force to one of the two particles, say particle 1, the system potential becomes $U(i_1,i_2)-f_1 a\cdot i_1$, where $U(i_1,i_2)$ is given by eq.~(\ref{U:def}). The transition rates (\ref{db1})-(\ref{db2}) are modified accordingly, and 
 one can evaluate the delivered power, as given by the product of the particle average velocity and the applied force $P_{out}=-a f_1 J_1$, and the average heat rate from the hot reservoir, eqs.(\ref{q:disc1})-(\ref{q:disc2}). In the presence of the force, one does not obtain a compact expression for the particle and heat currents however, the master equation for the steady state $P (i_1 , i_2 )$ can be easily solved. The efficiency $\eta=P_{out}/\langle{\dot Q_H}\rangle $ and the output power are shown in figure~\ref{fig:disc}: we find efficiency values  which are higher than those obtained for the continuous model, although the largest values of the efficiency ($\sim 50\%$)  are obtained for large coupling constant $k$ but close to the stall condition, where both $P_{out}$ and $\langle{\dot Q_H}\rangle$ vanish. In the region where the delivered power is maximum, the efficiency is $\sim 20\%$. Such a difference in the efficiency between the continuous and the discrete models can be understood if one notices that the continuous model does not exhibit a strong coupling between the input and the output energy currents. During its dynamic evolution the continuous model can exhibit time lapses where the center-of-mass coordinate does not advance, fluctuating back and forth. This is accompanied by simultaneous fluctuations of the relative coordinate $y$, leading to heat flowing between the two reservoirs. A typical example of such trajectories can occur close to the stall condition: if one  applies an external force on the system which is large enough to stall the motor, the system mean velocity, and thus the extracted power will become small, but the heat current between the two reservoirs will not vanish. 
 On the contrary  the discrete model exhibits  a stronger coupling between the input and the output cycles: when the particle on which the force is applied advances in the positive direction, there is a simultaneous contribution to the extracted work and to the heat current, and this results in higher values for the efficiency and the efficiency at maximum power \cite{Seifert2011a}.} 

{\it Conclusions}

In conclusion, we have shown that a periodic system, consisting of two Brownian particles,  can exhibit direct transport, and behave as an autonomous heat engine when an external mechanical force is applied. In the large coupling regime, the model is equivalent to a single Brownian particle in a position dependent temperature profile. However, the heat rates are well-defined quantities, given that each degree of freedom is in contact with its heat reservoir, and thus the efficiency of the heat engine can be evaluated for any value of the interaction strength.
We introduce a minimal discrete model that captures the essential features of the continuous Brownian motor, in particular the velocity scaling behaviour for small coupling, \alb{and that exhibits a larger efficiency that the continuous case, given the stronger coupling between the input and the output energy currents}.
 Finally, we emphasize that the model engine we propose is feasible of experimental realization by using, e.g., the setups considered in \cite{Ciliberto2013,Ciliberto2013a} or in \cite{Mehl2012}. \alb{The latter apparatus seems to be particularly suitable to test our findings as it consists of two paramagnetic colloidal particles sitting in effective sinusoidal potentials created by  toroidal laser traps.}

\begin{acknowledgments}
A.I. was supported by the Danish Council for Independent Research and the Villum Fonden.
\end{acknowledgments}

\def\url#1{}
\bibliography{bibliography}

\begin{thebibliography}{38}%
\makeatletter
\providecommand \@ifxundefined [1]{%
 \@ifx{#1\undefined}
}%
\providecommand \@ifnum [1]{%
 \ifnum #1\expandafter \@firstoftwo
 \else \expandafter \@secondoftwo
 \fi
}%
\providecommand \@ifx [1]{%
 \ifx #1\expandafter \@firstoftwo
 \else \expandafter \@secondoftwo
 \fi
}%
\providecommand \natexlab [1]{#1}%
\providecommand \enquote  [1]{``#1''}%
\providecommand \bibnamefont  [1]{#1}%
\providecommand \bibfnamefont [1]{#1}%
\providecommand \citenamefont [1]{#1}%
\providecommand \href@noop [0]{\@secondoftwo}%
\providecommand \href [0]{\begingroup \@sanitize@url \@href}%
\providecommand \@href[1]{\@@startlink{#1}\@@href}%
\providecommand \@@href[1]{\endgroup#1\@@endlink}%
\providecommand \@sanitize@url [0]{\catcode `\\12\catcode `\$12\catcode
  `\&12\catcode `\#12\catcode `\^12\catcode `\_12\catcode `\%12\relax}%
\providecommand \@@startlink[1]{}%
\providecommand \@@endlink[0]{}%
\providecommand \url  [0]{\begingroup\@sanitize@url \@url }%
\providecommand \@url [1]{\endgroup\@href {#1}{\urlprefix }}%
\providecommand \urlprefix  [0]{URL }%
\providecommand \Eprint [0]{\href }%
\providecommand \doibase [0]{http://dx.doi.org/}%
\providecommand \selectlanguage [0]{\@gobble}%
\providecommand \bibinfo  [0]{\@secondoftwo}%
\providecommand \bibfield  [0]{\@secondoftwo}%
\providecommand \translation [1]{[#1]}%
\providecommand \BibitemOpen [0]{}%
\providecommand \bibitemStop [0]{}%
\providecommand \bibitemNoStop [0]{.\EOS\space}%
\providecommand \EOS [0]{\spacefactor3000\relax}%
\providecommand \BibitemShut  [1]{\csname bibitem#1\endcsname}%
\let\auto@bib@innerbib\@empty
%</preamble>
\bibitem [{\citenamefont {Landauer}(1988)}]{Landauer1988}%
  \BibitemOpen
  \bibfield  {author} {\bibinfo {author} {\bibfnamefont {R.}~\bibnamefont
  {Landauer}},\ }\href {\doibase 10.1007/BF01011555} {\bibfield  {journal}
  {\bibinfo  {journal} {Journal of Statistical Physics}\ }\textbf {\bibinfo
  {volume} {53}},\ \bibinfo {pages} {233} (\bibinfo {year} {1988})}\BibitemShut
  {NoStop}%
\bibitem [{\citenamefont {Hondou}\ and\ \citenamefont
  {Sekimoto}(2000)}]{Hondou2000}%
  \BibitemOpen
  \bibfield  {author} {\bibinfo {author} {\bibfnamefont {T.}~\bibnamefont
  {Hondou}}\ and\ \bibinfo {author} {\bibfnamefont {K.}~\bibnamefont
  {Sekimoto}},\ }\href {\doibase 10.1103/PhysRevE.62.6021} {\bibfield
  {journal} {\bibinfo  {journal} {Phys. Rev. E}\ }\textbf {\bibinfo {volume}
  {62}},\ \bibinfo {pages} {6021} (\bibinfo {year} {2000})}\BibitemShut
  {NoStop}%
\bibitem [{\citenamefont {Benjamin}\ and\ \citenamefont
  {Kawai}(2008)}]{Benjamin08}%
  \BibitemOpen
  \bibfield  {author} {\bibinfo {author} {\bibfnamefont {R.}~\bibnamefont
  {Benjamin}}\ and\ \bibinfo {author} {\bibfnamefont {R.}~\bibnamefont
  {Kawai}},\ }\href {\doibase 10.1103/PhysRevE.77.051132} {\bibfield  {journal}
  {\bibinfo  {journal} {Phys. Rev. E}\ }\textbf {\bibinfo {volume} {77}},\
  \bibinfo {pages} {051132} (\bibinfo {year} {2008})}\BibitemShut {NoStop}%
\bibitem [{\citenamefont {Kay}\ \emph {et~al.}(2007)\citenamefont {Kay},
  \citenamefont {Leigh},\ and\ \citenamefont {Zerbetto}}]{Kay2007}%
  \BibitemOpen
  \bibfield  {author} {\bibinfo {author} {\bibfnamefont {E.~R.}\ \bibnamefont
  {Kay}}, \bibinfo {author} {\bibfnamefont {D.~A.}\ \bibnamefont {Leigh}}, \
  and\ \bibinfo {author} {\bibfnamefont {F.}~\bibnamefont {Zerbetto}},\ }\href
  {\doibase 10.1002/anie.200504313} {\bibfield  {journal} {\bibinfo  {journal}
  {Angew. Chem. Int. Ed.}\ }\textbf {\bibinfo {volume} {46}},\ \bibinfo {pages}
  {72} (\bibinfo {year} {2007})}\BibitemShut {NoStop}%
\bibitem [{\citenamefont {Liu}\ and\ \citenamefont {Liu}(2009)}]{Liu2009}%
  \BibitemOpen
  \bibfield  {author} {\bibinfo {author} {\bibfnamefont {H.}~\bibnamefont
  {Liu}}\ and\ \bibinfo {author} {\bibfnamefont {D.}~\bibnamefont {Liu}},\
  }\href {\doibase 10.1039/B822719E} {\bibfield  {journal} {\bibinfo  {journal}
  {Chem. Commun.}\ ,\ \bibinfo {pages} {2625}} (\bibinfo {year}
  {2009})}\BibitemShut {NoStop}%
\bibitem [{\citenamefont {Lund}\ \emph {et~al.}(2010)\citenamefont {Lund},
  \citenamefont {Manzo}, \citenamefont {Dabby}, \citenamefont {Michelotti},
  \citenamefont {Johnson-Buck}, \citenamefont {Nangreave}, \citenamefont
  {Taylor}, \citenamefont {Pei}, \citenamefont {Stojanovic}, \citenamefont
  {Walter}, \citenamefont {Winfree},\ and\ \citenamefont {Yan}}]{Lund2010}%
  \BibitemOpen
  \bibfield  {author} {\bibinfo {author} {\bibfnamefont {K.}~\bibnamefont
  {Lund}}, \bibinfo {author} {\bibfnamefont {A.~J.}\ \bibnamefont {Manzo}},
  \bibinfo {author} {\bibfnamefont {N.}~\bibnamefont {Dabby}}, \bibinfo
  {author} {\bibfnamefont {N.}~\bibnamefont {Michelotti}}, \bibinfo {author}
  {\bibfnamefont {A.}~\bibnamefont {Johnson-Buck}}, \bibinfo {author}
  {\bibfnamefont {J.}~\bibnamefont {Nangreave}}, \bibinfo {author}
  {\bibfnamefont {S.}~\bibnamefont {Taylor}}, \bibinfo {author} {\bibfnamefont
  {R.}~\bibnamefont {Pei}}, \bibinfo {author} {\bibfnamefont {M.~N.}\
  \bibnamefont {Stojanovic}}, \bibinfo {author} {\bibfnamefont {N.~G.}\
  \bibnamefont {Walter}}, \bibinfo {author} {\bibfnamefont {E.}~\bibnamefont
  {Winfree}}, \ and\ \bibinfo {author} {\bibfnamefont {H.}~\bibnamefont
  {Yan}},\ }\href {\doibase doi:10.1038/nature09012} {\bibfield  {journal}
  {\bibinfo  {journal} {Nature}\ }\textbf {\bibinfo {volume} {465}},\ \bibinfo
  {pages} {206} (\bibinfo {year} {2010})}\BibitemShut {NoStop}%
\bibitem [{\citenamefont {Reimann}(2002)}]{Reimann02}%
  \BibitemOpen
  \bibfield  {author} {\bibinfo {author} {\bibfnamefont {P.}~\bibnamefont
  {Reimann}},\ }\href {\doibase
  http://dx.doi.org/10.1016/S0370-1573(01)00081-3} {\bibfield  {journal}
  {\bibinfo  {journal} {Physics Reports}\ }\textbf {\bibinfo {volume} {361}},\
  \bibinfo {pages} {57 } (\bibinfo {year} {2002})}\BibitemShut {NoStop}%
\bibitem [{\citenamefont {van~den Broek}\ \emph {et~al.}(2009)\citenamefont
  {van~den Broek}, \citenamefont {Eichhorn},\ and\ \citenamefont {den
  Broeck}}]{VandenBroeck08}%
  \BibitemOpen
  \bibfield  {author} {\bibinfo {author} {\bibfnamefont {M.}~\bibnamefont
  {van~den Broek}}, \bibinfo {author} {\bibfnamefont {R.}~\bibnamefont
  {Eichhorn}}, \ and\ \bibinfo {author} {\bibfnamefont {C.~V.}\ \bibnamefont
  {den Broeck}},\ }\href {http://stacks.iop.org/0295-5075/86/i=3/a=30002}
  {\bibfield  {journal} {\bibinfo  {journal} {EPL (Europhysics Letters)}\
  }\textbf {\bibinfo {volume} {86}},\ \bibinfo {pages} {30002} (\bibinfo {year}
  {2009})}\BibitemShut {NoStop}%
\bibitem [{\citenamefont {von Gehlen}\ \emph {et~al.}(2009)\citenamefont {von
  Gehlen}, \citenamefont {Evstigneev},\ and\ \citenamefont
  {Reimann}}]{Gehlen09}%
  \BibitemOpen
  \bibfield  {author} {\bibinfo {author} {\bibfnamefont {S.}~\bibnamefont {von
  Gehlen}}, \bibinfo {author} {\bibfnamefont {M.}~\bibnamefont {Evstigneev}}, \
  and\ \bibinfo {author} {\bibfnamefont {P.}~\bibnamefont {Reimann}},\ }\href
  {\doibase 10.1103/PhysRevE.79.031114} {\bibfield  {journal} {\bibinfo
  {journal} {Phys. Rev. E}\ }\textbf {\bibinfo {volume} {79}},\ \bibinfo
  {pages} {031114} (\bibinfo {year} {2009})}\BibitemShut {NoStop}%
\bibitem [{\citenamefont {Schmiedl}\ and\ \citenamefont
  {Seifert}(2008)}]{Schmiedl08}%
  \BibitemOpen
  \bibfield  {author} {\bibinfo {author} {\bibfnamefont {T.}~\bibnamefont
  {Schmiedl}}\ and\ \bibinfo {author} {\bibfnamefont {U.}~\bibnamefont
  {Seifert}},\ }\href {http://stacks.iop.org/0295-5075/81/i=2/a=20003}
  {\bibfield  {journal} {\bibinfo  {journal} {EPL (Europhysics Letters)}\
  }\textbf {\bibinfo {volume} {81}},\ \bibinfo {pages} {20003} (\bibinfo {year}
  {2008})}\BibitemShut {NoStop}%
\bibitem [{\citenamefont {Brandner}\ \emph {et~al.}(2015)\citenamefont
  {Brandner}, \citenamefont {Saito},\ and\ \citenamefont
  {Seifert}}]{Brandner15}%
  \BibitemOpen
  \bibfield  {author} {\bibinfo {author} {\bibfnamefont {K.}~\bibnamefont
  {Brandner}}, \bibinfo {author} {\bibfnamefont {K.}~\bibnamefont {Saito}}, \
  and\ \bibinfo {author} {\bibfnamefont {U.}~\bibnamefont {Seifert}},\ }\href
  {\doibase 10.1103/PhysRevX.5.031019} {\bibfield  {journal} {\bibinfo
  {journal} {Phys. Rev. X}\ }\textbf {\bibinfo {volume} {5}},\ \bibinfo {pages}
  {031019} (\bibinfo {year} {2015})}\BibitemShut {NoStop}%
\bibitem [{\citenamefont {Blickle}\ and\ \citenamefont
  {Bechinger}(2012)}]{Blickle2012}%
  \BibitemOpen
  \bibfield  {author} {\bibinfo {author} {\bibfnamefont {V.}~\bibnamefont
  {Blickle}}\ and\ \bibinfo {author} {\bibfnamefont {C.}~\bibnamefont
  {Bechinger}},\ }\href {\doibase 10.1038/nphys2163} {\bibfield  {journal}
  {\bibinfo  {journal} {Nat Phys}\ }\textbf {\bibinfo {volume} {8}},\ \bibinfo
  {pages} {143} (\bibinfo {year} {2012})}\BibitemShut {NoStop}%
\bibitem [{\citenamefont {Matsuo}\ and\ \citenamefont {ichi
  Sasa}(2000)}]{Matsuo2000}%
  \BibitemOpen
  \bibfield  {author} {\bibinfo {author} {\bibfnamefont {M.}~\bibnamefont
  {Matsuo}}\ and\ \bibinfo {author} {\bibfnamefont {S.}~\bibnamefont {ichi
  Sasa}},\ }\href {\doibase http://dx.doi.org/10.1016/S0378-4371(99)00365-9}
  {\bibfield  {journal} {\bibinfo  {journal} {Physica A: Statistical Mechanics
  and its Applications}\ }\textbf {\bibinfo {volume} {276}},\ \bibinfo {pages}
  {188 } (\bibinfo {year} {2000})}\BibitemShut {NoStop}%
\bibitem [{\citenamefont {Berger}\ \emph {et~al.}(2009)\citenamefont {Berger},
  \citenamefont {Schmiedl},\ and\ \citenamefont {Seifert}}]{Berger09}%
  \BibitemOpen
  \bibfield  {author} {\bibinfo {author} {\bibfnamefont {F.}~\bibnamefont
  {Berger}}, \bibinfo {author} {\bibfnamefont {T.}~\bibnamefont {Schmiedl}}, \
  and\ \bibinfo {author} {\bibfnamefont {U.}~\bibnamefont {Seifert}},\ }\href
  {\doibase 10.1103/PhysRevE.79.031118} {\bibfield  {journal} {\bibinfo
  {journal} {Phys. Rev. E}\ }\textbf {\bibinfo {volume} {79}},\ \bibinfo
  {pages} {031118} (\bibinfo {year} {2009})}\BibitemShut {NoStop}%
\bibitem [{\citenamefont {Feynman}\ \emph {et~al.}(1963)\citenamefont
  {Feynman}, \citenamefont {Leighton},\ and\ \citenamefont
  {Sands}}]{Feynman:1963}%
  \BibitemOpen
  \bibfield  {author} {\bibinfo {author} {\bibfnamefont {R.~P. R.~P.}\
  \bibnamefont {Feynman}}, \bibinfo {author} {\bibfnamefont {R.~B.}\
  \bibnamefont {Leighton}}, \ and\ \bibinfo {author} {\bibfnamefont {M.~L.
  M.~L.}\ \bibnamefont {Sands}},\ }\href@noop {} {\emph {\bibinfo {title} {The
  {Feynman} lectures on physics}}}\ (\bibinfo  {publisher} {Addison-Wesley},\
  \bibinfo {address} {Reading, MA},\ \bibinfo {year} {1963})\ pp.\ \bibinfo
  {pages} {xii + 513},\ \bibinfo {note} {three volumes.}\BibitemShut {Stop}%
\bibitem [{\citenamefont {Van~den Broeck}\ \emph {et~al.}(2004)\citenamefont
  {Van~den Broeck}, \citenamefont {Kawai},\ and\ \citenamefont
  {Meurs}}]{VandenBroeck04}%
  \BibitemOpen
  \bibfield  {author} {\bibinfo {author} {\bibfnamefont {C.}~\bibnamefont
  {Van~den Broeck}}, \bibinfo {author} {\bibfnamefont {R.}~\bibnamefont
  {Kawai}}, \ and\ \bibinfo {author} {\bibfnamefont {P.}~\bibnamefont
  {Meurs}},\ }\href {\doibase 10.1103/PhysRevLett.93.090601} {\bibfield
  {journal} {\bibinfo  {journal} {Phys. Rev. Lett.}\ }\textbf {\bibinfo
  {volume} {93}},\ \bibinfo {pages} {090601} (\bibinfo {year}
  {2004})}\BibitemShut {NoStop}%
\bibitem [{\citenamefont {Fruleux}\ \emph {et~al.}(2012)\citenamefont
  {Fruleux}, \citenamefont {Kawai},\ and\ \citenamefont
  {Sekimoto}}]{Fruleux12}%
  \BibitemOpen
  \bibfield  {author} {\bibinfo {author} {\bibfnamefont {A.}~\bibnamefont
  {Fruleux}}, \bibinfo {author} {\bibfnamefont {R.}~\bibnamefont {Kawai}}, \
  and\ \bibinfo {author} {\bibfnamefont {K.}~\bibnamefont {Sekimoto}},\ }\href
  {\doibase 10.1103/PhysRevLett.108.160601} {\bibfield  {journal} {\bibinfo
  {journal} {Phys. Rev. Lett.}\ }\textbf {\bibinfo {volume} {108}},\ \bibinfo
  {pages} {160601} (\bibinfo {year} {2012})}\BibitemShut {NoStop}%
\bibitem [{\citenamefont {Holubec}\ \emph {et~al.}(2017)\citenamefont
  {Holubec}, \citenamefont {Ryabov}, \citenamefont {Yaghoubi}, \citenamefont
  {Varga}, \citenamefont {Khodaee}, \citenamefont {Foulaadvand},\ and\
  \citenamefont {Chvosta}}]{Holubec17}%
  \BibitemOpen
  \bibfield  {author} {\bibinfo {author} {\bibfnamefont {V.}~\bibnamefont
  {Holubec}}, \bibinfo {author} {\bibfnamefont {A.}~\bibnamefont {Ryabov}},
  \bibinfo {author} {\bibfnamefont {M.~H.}\ \bibnamefont {Yaghoubi}}, \bibinfo
  {author} {\bibfnamefont {M.}~\bibnamefont {Varga}}, \bibinfo {author}
  {\bibfnamefont {A.}~\bibnamefont {Khodaee}}, \bibinfo {author} {\bibfnamefont
  {M.~E.}\ \bibnamefont {Foulaadvand}}, \ and\ \bibinfo {author} {\bibfnamefont
  {P.}~\bibnamefont {Chvosta}},\ }\href {\doibase 10.3390/e19040119} {\bibfield
   {journal} {\bibinfo  {journal} {Entropy}\ }\textbf {\bibinfo {volume} {19}}
  (\bibinfo {year} {2017}),\ 10.3390/e19040119}\BibitemShut {NoStop}%
\bibitem [{\citenamefont {Filliger}\ and\ \citenamefont
  {Reimann}(2007)}]{Reimann07}%
  \BibitemOpen
  \bibfield  {author} {\bibinfo {author} {\bibfnamefont {R.}~\bibnamefont
  {Filliger}}\ and\ \bibinfo {author} {\bibfnamefont {P.}~\bibnamefont
  {Reimann}},\ }\href {\doibase 10.1103/PhysRevLett.99.230602} {\bibfield
  {journal} {\bibinfo  {journal} {Phys. Rev. Lett.}\ }\textbf {\bibinfo
  {volume} {99}},\ \bibinfo {pages} {230602} (\bibinfo {year}
  {2007})}\BibitemShut {NoStop}%
\bibitem [{\citenamefont {Dotsenko}\ \emph {et~al.}(2013)\citenamefont
  {Dotsenko}, \citenamefont {Macio\l{}ek}, \citenamefont {Vasilyev},\ and\
  \citenamefont {Oshanin}}]{Dotsenko13}%
  \BibitemOpen
  \bibfield  {author} {\bibinfo {author} {\bibfnamefont {V.}~\bibnamefont
  {Dotsenko}}, \bibinfo {author} {\bibfnamefont {A.}~\bibnamefont
  {Macio\l{}ek}}, \bibinfo {author} {\bibfnamefont {O.}~\bibnamefont
  {Vasilyev}}, \ and\ \bibinfo {author} {\bibfnamefont {G.}~\bibnamefont
  {Oshanin}},\ }\href {\doibase 10.1103/PhysRevE.87.062130} {\bibfield
  {journal} {\bibinfo  {journal} {Phys. Rev. E}\ }\textbf {\bibinfo {volume}
  {87}},\ \bibinfo {pages} {062130} (\bibinfo {year} {2013})}\BibitemShut
  {NoStop}%
\bibitem [{Arg()}]{Argun17}%
  \BibitemOpen
  \href@noop {} {}\bibinfo {note} {Aykut Argun, Jalpa Soni, Lennart Dabelow,
  Stefano Bo, Giuseppe Pesce, Ralf Eichhorn, Giovanni Volpe, Experimental
  Realization of a Minimal Microscopic Heat Engine ,
  arXiv:1708.07197}\BibitemShut {NoStop}%
\bibitem [{\citenamefont {Gomez-Marin}\ and\ \citenamefont
  {Sancho}(2005)}]{Sancho05}%
  \BibitemOpen
  \bibfield  {author} {\bibinfo {author} {\bibfnamefont {A.}~\bibnamefont
  {Gomez-Marin}}\ and\ \bibinfo {author} {\bibfnamefont {J.~M.}\ \bibnamefont
  {Sancho}},\ }\href {\doibase 10.1103/PhysRevE.71.021101} {\bibfield
  {journal} {\bibinfo  {journal} {Phys. Rev. E}\ }\textbf {\bibinfo {volume}
  {71}},\ \bibinfo {pages} {021101} (\bibinfo {year} {2005})}\BibitemShut
  {NoStop}%
\bibitem [{\citenamefont {Sekimoto}(2010)}]{Sekimoto:Book}%
  \BibitemOpen
  \bibfield  {author} {\bibinfo {author} {\bibfnamefont {K.}~\bibnamefont
  {Sekimoto}},\ }\href {\doibase 10.1007/978-3-642-05411-2} {\emph {\bibinfo
  {title} {Stochastic Energetics}}},\ Lecture Notes in Physics\ (\bibinfo
  {publisher} {Springer},\ \bibinfo {year} {2010})\BibitemShut {NoStop}%
\bibitem [{\citenamefont {Imparato}\ \emph {et~al.}(2007)\citenamefont
  {Imparato}, \citenamefont {Peliti}, \citenamefont {Pesce}, \citenamefont
  {Rusciano},\ and\ \citenamefont {Sasso}}]{Imparato07}%
  \BibitemOpen
  \bibfield  {author} {\bibinfo {author} {\bibfnamefont {A.}~\bibnamefont
  {Imparato}}, \bibinfo {author} {\bibfnamefont {L.}~\bibnamefont {Peliti}},
  \bibinfo {author} {\bibfnamefont {G.}~\bibnamefont {Pesce}}, \bibinfo
  {author} {\bibfnamefont {G.}~\bibnamefont {Rusciano}}, \ and\ \bibinfo
  {author} {\bibfnamefont {A.}~\bibnamefont {Sasso}},\ }\href {\doibase
  10.1103/PhysRevE.76.050101} {\bibfield  {journal} {\bibinfo  {journal} {Phys.
  Rev. E}\ }\textbf {\bibinfo {volume} {76}},\ \bibinfo {pages} {050101}
  (\bibinfo {year} {2007})}\BibitemShut {NoStop}%
\bibitem [{\citenamefont {Fogedby}\ and\ \citenamefont
  {Imparato}(2012)}]{Fogedby12}%
  \BibitemOpen
  \bibfield  {author} {\bibinfo {author} {\bibfnamefont {H.~C.}\ \bibnamefont
  {Fogedby}}\ and\ \bibinfo {author} {\bibfnamefont {A.}~\bibnamefont
  {Imparato}},\ }\href {http://stacks.iop.org/1742-5468/2012/i=04/a=P04005}
  {\bibfield  {journal} {\bibinfo  {journal} {Journal of Statistical Mechanics:
  Theory and Experiment}\ }\textbf {\bibinfo {volume} {2012}},\ \bibinfo
  {pages} {P04005} (\bibinfo {year} {2012})}\BibitemShut {NoStop}%
\bibitem [{\citenamefont {Fogedby}\ and\ \citenamefont
  {Imparato}(2014)}]{Fogedby14}%
  \BibitemOpen
  \bibfield  {author} {\bibinfo {author} {\bibfnamefont {H.~C.}\ \bibnamefont
  {Fogedby}}\ and\ \bibinfo {author} {\bibfnamefont {A.}~\bibnamefont
  {Imparato}},\ }\href {http://stacks.iop.org/1742-5468/2014/i=11/a=P11011}
  {\bibfield  {journal} {\bibinfo  {journal} {Journal of Statistical Mechanics:
  Theory and Experiment}\ }\textbf {\bibinfo {volume} {2014}},\ \bibinfo
  {pages} {P11011} (\bibinfo {year} {2014})}\BibitemShut {NoStop}%
\bibitem [{\citenamefont {Risken}(1996)}]{Risken}%
  \BibitemOpen
  \bibfield  {author} {\bibinfo {author} {\bibfnamefont {H.}~\bibnamefont
  {Risken}},\ }\href {https://books.google.dk/books?id=MG2V9vTgSgEC} {\emph
  {\bibinfo {title} {The Fokker-Planck Equation: Methods of Solution and
  Applications}}},\ Springer Series in Synergetics\ (\bibinfo  {publisher}
  {Springer Berlin Heidelberg},\ \bibinfo {year} {1996})\BibitemShut {NoStop}%
\bibitem [{\citenamefont {Golubeva}\ \emph {et~al.}(2012)\citenamefont
  {Golubeva}, \citenamefont {Imparato},\ and\ \citenamefont
  {Peliti}}]{Golubeva2012}%
  \BibitemOpen
  \bibfield  {author} {\bibinfo {author} {\bibfnamefont {N.}~\bibnamefont
  {Golubeva}}, \bibinfo {author} {\bibfnamefont {A.}~\bibnamefont {Imparato}},
  \ and\ \bibinfo {author} {\bibfnamefont {L.}~\bibnamefont {Peliti}},\
  }\href@noop {} {\bibfield  {journal} {\bibinfo  {journal} {Europhys. Lett.}\
  }\textbf {\bibinfo {volume} {97}},\ \bibinfo {pages} {60005} (\bibinfo {year}
  {2012})}\BibitemShut {NoStop}%
\bibitem [{\citenamefont {Van~den Broeck}\ \emph {et~al.}(2012)\citenamefont
  {Van~den Broeck}, \citenamefont {Kumar},\ and\ \citenamefont
  {Lindenberg}}]{VandenBroeck2012}%
  \BibitemOpen
  \bibfield  {author} {\bibinfo {author} {\bibfnamefont {C.}~\bibnamefont
  {Van~den Broeck}}, \bibinfo {author} {\bibfnamefont {N.}~\bibnamefont
  {Kumar}}, \ and\ \bibinfo {author} {\bibfnamefont {K.}~\bibnamefont
  {Lindenberg}},\ }\href@noop {} {\bibfield  {journal} {\bibinfo  {journal}
  {Phys. Rev. Lett.}\ }\textbf {\bibinfo {volume} {108}},\ \bibinfo {pages}
  {210602} (\bibinfo {year} {2012})}\BibitemShut {NoStop}%
\bibitem [{\citenamefont {Chaikin}\ and\ \citenamefont
  {Lubensky}(1995)}]{Chaikin}%
  \BibitemOpen
  \bibfield  {author} {\bibinfo {author} {\bibfnamefont {P.~M.}\ \bibnamefont
  {Chaikin}}\ and\ \bibinfo {author} {\bibfnamefont {T.~C.}\ \bibnamefont
  {Lubensky}},\ }\href@noop {} {\emph {\bibinfo {title} {Principles of
  Condensed Matter Physics}}}\ (\bibinfo  {publisher} {Cambridge University
  Press},\ \bibinfo {address} {Cambridge},\ \bibinfo {year} {1995})\BibitemShut
  {NoStop}%
\bibitem [{\citenamefont {Glauber}(1963)}]{Glauber63}%
  \BibitemOpen
  \bibfield  {author} {\bibinfo {author} {\bibfnamefont {R.~J.}\ \bibnamefont
  {Glauber}},\ }\href {\doibase 10.1063/1.1703954} {\bibfield  {journal}
  {\bibinfo  {journal} {Journal of Mathematical Physics}\ }\textbf {\bibinfo
  {volume} {4}},\ \bibinfo {pages} {294} (\bibinfo {year} {1963})},\ \Eprint
  {http://arxiv.org/abs/http://dx.doi.org/10.1063/1.1703954}
  {http://dx.doi.org/10.1063/1.1703954} \BibitemShut {NoStop}%
\bibitem [{\citenamefont {Lebowitz}\ and\ \citenamefont
  {Spohn}(1999)}]{Lebowitz1999}%
  \BibitemOpen
  \bibfield  {author} {\bibinfo {author} {\bibfnamefont {J.~L.}\ \bibnamefont
  {Lebowitz}}\ and\ \bibinfo {author} {\bibfnamefont {H.}~\bibnamefont
  {Spohn}},\ }\href {\doibase 10.1023/A:1004589714161} {\bibfield  {journal}
  {\bibinfo  {journal} {Journal of Statistical Physics}\ }\textbf {\bibinfo
  {volume} {95}},\ \bibinfo {pages} {333} (\bibinfo {year} {1999})}\BibitemShut
  {NoStop}%
\bibitem [{\citenamefont {Imparato}\ and\ \citenamefont
  {Peliti}(2007)}]{Imparato7a}%
  \BibitemOpen
  \bibfield  {author} {\bibinfo {author} {\bibfnamefont {A.}~\bibnamefont
  {Imparato}}\ and\ \bibinfo {author} {\bibfnamefont {L.}~\bibnamefont
  {Peliti}},\ }\href {http://stacks.iop.org/1742-5468/2007/i=02/a=L02001}
  {\bibfield  {journal} {\bibinfo  {journal} {Journal of Statistical Mechanics:
  Theory and Experiment}\ }\textbf {\bibinfo {volume} {2007}},\ \bibinfo
  {pages} {L02001} (\bibinfo {year} {2007})}\BibitemShut {NoStop}%
\bibitem [{\citenamefont {Seifert}(2012)}]{Seifert2012}%
  \BibitemOpen
  \bibfield  {author} {\bibinfo {author} {\bibfnamefont {U.}~\bibnamefont
  {Seifert}},\ }\href {http://stacks.iop.org/0034-4885/75/i=12/a=126001}
  {\bibfield  {journal} {\bibinfo  {journal} {Reports on Progress in Physics}\
  }\textbf {\bibinfo {volume} {75}},\ \bibinfo {pages} {126001} (\bibinfo
  {year} {2012})}\BibitemShut {NoStop}%
\bibitem [{\citenamefont {Seifert}(2011)}]{Seifert2011a}%
  \BibitemOpen
  \bibfield  {author} {\bibinfo {author} {\bibfnamefont {U.}~\bibnamefont
  {Seifert}},\ }\href@noop {} {\bibfield  {journal} {\bibinfo  {journal} {Phys.
  Rev. Lett.}\ }\textbf {\bibinfo {volume} {106}},\ \bibinfo {pages} {020601}
  (\bibinfo {year} {2011})}\BibitemShut {NoStop}%
\bibitem [{\citenamefont {Ciliberto}\ \emph
  {et~al.}(2013{\natexlab{a}})\citenamefont {Ciliberto}, \citenamefont
  {Imparato}, \citenamefont {Naert},\ and\ \citenamefont
  {Tanase}}]{Ciliberto2013}%
  \BibitemOpen
  \bibfield  {author} {\bibinfo {author} {\bibfnamefont {S.}~\bibnamefont
  {Ciliberto}}, \bibinfo {author} {\bibfnamefont {A.}~\bibnamefont {Imparato}},
  \bibinfo {author} {\bibfnamefont {A.}~\bibnamefont {Naert}}, \ and\ \bibinfo
  {author} {\bibfnamefont {M.}~\bibnamefont {Tanase}},\ }\href {\doibase
  10.1103/PhysRevLett.110.180601} {\bibfield  {journal} {\bibinfo  {journal}
  {Phys. Rev. Lett.}\ }\textbf {\bibinfo {volume} {110}},\ \bibinfo {pages}
  {180601} (\bibinfo {year} {2013}{\natexlab{a}})}\BibitemShut {NoStop}%
\bibitem [{\citenamefont {Ciliberto}\ \emph
  {et~al.}(2013{\natexlab{b}})\citenamefont {Ciliberto}, \citenamefont
  {Imparato}, \citenamefont {Naert},\ and\ \citenamefont
  {Tanase}}]{Ciliberto2013a}%
  \BibitemOpen
  \bibfield  {author} {\bibinfo {author} {\bibfnamefont {S.}~\bibnamefont
  {Ciliberto}}, \bibinfo {author} {\bibfnamefont {A.}~\bibnamefont {Imparato}},
  \bibinfo {author} {\bibfnamefont {A.}~\bibnamefont {Naert}}, \ and\ \bibinfo
  {author} {\bibfnamefont {M.}~\bibnamefont {Tanase}},\ }\href
  {http://stacks.iop.org/1742-5468/2013/i=12/a=P12014} {\bibfield  {journal}
  {\bibinfo  {journal} {Journal of Statistical Mechanics: Theory and
  Experiment}\ }\textbf {\bibinfo {volume} {2013}},\ \bibinfo {pages} {P12014}
  (\bibinfo {year} {2013}{\natexlab{b}})}\BibitemShut {NoStop}%
\bibitem [{\citenamefont {Mehl}\ \emph {et~al.}(2012)\citenamefont {Mehl},
  \citenamefont {Lander}, \citenamefont {Bechinger}, \citenamefont {Blickle},\
  and\ \citenamefont {Seifert}}]{Mehl2012}%
  \BibitemOpen
  \bibfield  {author} {\bibinfo {author} {\bibfnamefont {J.}~\bibnamefont
  {Mehl}}, \bibinfo {author} {\bibfnamefont {B.}~\bibnamefont {Lander}},
  \bibinfo {author} {\bibfnamefont {C.}~\bibnamefont {Bechinger}}, \bibinfo
  {author} {\bibfnamefont {V.}~\bibnamefont {Blickle}}, \ and\ \bibinfo
  {author} {\bibfnamefont {U.}~\bibnamefont {Seifert}},\ }\href {\doibase
  10.1103/PhysRevLett.108.220601} {\bibfield  {journal} {\bibinfo  {journal}
  {Phys. Rev. Lett.}\ }\textbf {\bibinfo {volume} {108}},\ \bibinfo {pages}
  {220601} (\bibinfo {year} {2012})}\BibitemShut {NoStop}%
\end{thebibliography}%
\newpage

\appendix
\section{Heat rates}
\label{app2}
Following the standard approach in stochastic thermodynamics \cite{Sekimoto:Book}, we define the rate of heat exchanged by particle $i$ with the corresponding heat bath as the rate of work performed by the heat bath on the particle,
\begin{equation}
\dot Q_i= \dot x_i(-\dot x_i+\eta_i)= \pq{-\partial_i V(x_1,x_2)+\eta_i}\partial_i V(x_1,x_2),
\end{equation} 
using the Stratonovich form of stochastic calculus.

By introducing the joint probability distribution $\Phi(x_1,x_2,Q_i,t)$ and using a standard approach \cite{Risken}, one 
obtains the Fokker-Planck equation for $\Phi(x_1,x_2,Q_i,t)$,
%\begin{eqnarray}
%&&\partial_t\Phi(x_1,x_2,Q_i,t)=\nonumber \\
%&&\frac{\partial}{\partial Q_i} \pq{\p{\partial_i V}^2 +T_i \partial_{x_i}(\partial_{x_i} V) +T_i (\partial_{x_i} V) \partial_{x_i}+ T_1 %\frac{\partial}{\partial Q_i} }\Phi \nonumber .
%\end{eqnarray} 
\begin{eqnarray}
&&\partial_t\Phi=\partial_i\pq{\partial_iV+T_i\partial_i+T_i\frac{\partial}{\partial Q_i}\partial_iV}\Phi
\nonumber
\\
&&+\frac{\partial}{\partial Q_i} \pq{\p{\partial_i V}^2 +T_i \partial_iV\partial_i + 
T_i\frac{\partial}{\partial Q_i}(\partial_iV)^2}\Phi.
\nonumber
\\
\end{eqnarray}
The average heat rate is then given by
\begin{eqnarray}
\average{\dot Q_i}&=&\partial_t \int \D Q_i \D x_1\D x_2\,  Q_i \Phi(x_1,x_2,Q_i,t)\nonumber \\
&=&\average{T_i \partial_i^2V(x_1,x_2)-\p{\partial_i V(x_1,x_2)}^2}, 
\end{eqnarray} 
corresponding to eq.~(\ref{dotQ}) in the main text.
\section{Large coupling regime}
\label{app1}
We introduce the two variables $x=(x_1+x_2)/2$ and $y=(x_1-x_2)/2$, 
and notice that for large coupling $k$ the variable $y$ is suppressed allowing the expansions 
\begin{equation}
F_{1}(x_1)\simeq F_1(x)+F'_1(x) y, \quad F_{2}(x_2)\simeq F_2(x)-F'_2(x) y.
\end{equation} 
Furthermore, we assume that the interaction potential $u(z)$ has a minimum in $z=0$, so that $k u'(2 y)\simeq2 k y$.
By insertion we obtain
\begin{eqnarray}
\dot x &=&\frac 1 2 \pq{F_1(x)+F_2(x)}+\frac y 2 \pq{F'_1(x) -F'_2(x)}\nonumber\\
&&+\frac 1 2 (\eta_1+\eta_2) \label{eq:x}, \\
\dot y &=&\frac 1 2 \pq{F_1(x)-F_2(x)}+\frac y 2 \pq{F'_1(x) +F'_2(x)}-2 k y\nonumber\\
&& + \frac 1 2 (\eta_1-\eta_2).
\label{eq:y}
\end{eqnarray} 
Imposing $\dot y=0$, from eq.~(\ref{eq:y}) one finds 
\begin{equation}
y=\frac{ F_1(x)-F_2(x) +\eta_1-\eta_2}{4 k-F_1(x)+F_2(x)} 
\label{eq:y:adiab}
\end{equation} 
and by substituting eq.~(\ref{eq:y:adiab}) in eq.~(\ref{eq:x}) a single Langevin equation for the center of mass coordinate,
\begin{eqnarray}
\dot x&=&(F_1(x) +\eta_1(t))s_1(x)+(F_2(x)+\eta_2(t)) s_2(x)\nonumber\\
&=&h(x) +g(x) \xi(t),
\label{eq:x1}
\end{eqnarray} 
where 
\begin{eqnarray}
h(x)&=&F_1(x) s_1(x)+F_2(x) s_2(x),\\
s_1(x)&=&\frac{2 k -F'_2(x)}{4 k -(F'_1(x)+F'_2(x))},\\
s_2(x)&=&\frac{2 k -F'_1(x)}{4 k -(F'_1(x)+F'_2(x))},
\end{eqnarray} 
and
\begin{equation}
g(x)=\sqrt{T_1 s_1^2(x) +T_2 s_2^2(x)}.
\label{eq:g}
\end{equation} 
Here $\xi(t)$ is a white Gaussian noise with correlations $\average{\xi(t) \xi(t')}=2 \delta(t-t')$.
 
The Fokker--Planck equation corresponding to eq.~(\ref{eq:x1}) has the form  \cite{Risken}
\begin{equation}
\partial_t P(x,t)=\partial_x \pq{-h(x) + g(x)g'(x)+g^2(x)\partial_x} P(x,t),
\end{equation} 
with steady state solution 
\begin{equation}
P(x)=\frac{\E^{-U(x)}}{g(x)} \pq{c_1-c_2 I(x)},
\label{eq:pss}
\end{equation} 
where
\begin{equation}
U(x)=-\int_0^x \D y\, \frac{h(y)}{g^2(y)}, 
\label{eq:U:app}
\end{equation} 
and 
\begin{equation}
I(x)=\int_0^x \D y\,  \frac{\E^{U(y)}}{g(y)}.
\end{equation} 
Imposing the periodicity condition on the steady state  solution (\ref{eq:pss}),
i.e., $P(x)=P(x+L)$, where $L=\max(L_1,L_2)$, 
and noticing that $g(x)$ as defined in eq.~(\ref{eq:g}) is periodic, $g(0)=g(L)$, and that $U(0)=I(0)=0$, 
we obtain 
\begin{equation}
c_1=\E^{-U(L)}\pq{c_1-c_2 I(L)}
\label{eq:c1}
\end{equation} 
which solved for $c_1$
yields the distribution
\begin{eqnarray}
P(x)&=&c_2 \frac{\E^{-U(x)}}{g(x)} \p{\frac{I(L)}{1-\E^{U(L)}}- I(x)}\nonumber\\
&=& c_2 \frac{\E^{-U(x)}}{(1-\E^{\bar f L}) g(x)}\int_x^{x+L}  \frac{\E^{U(y)}}{g(y)}\,  d y, 
\label{eq:pss1}
\end{eqnarray} 
with $\bar f  = [U (x+L)-U (x)]/L$. The constant $c_2$ is fixed by the normalization condition $\int_0^L P(x)=1$.
Furthermore, it follows that $c_2$ is the steady state current
\begin{equation}
c_2=\Jss= \pq{h(x) - g(x)g'(x)-g^2(x)\partial_x} P(x),
\end{equation} 
yielding eq.~(\ref{cur2}) in the main text.
The $x$ coordinate steady state velocity is finally given by 
\begin{equation}
\bar v=L\cdot \Jss.
\label{eq:vx}
\end{equation} 

It is convenient to decompose the potentials $V_i(x)$ in their Fourier series
\begin{eqnarray}
V_1(x)&=&\sum_{q_1}\E^{\ii q_1 x}v_{1,q_1},\\
V_2(x)&=&\sum_{q_2}\E^{\ii q_2 x}v_{2,q_2},
\end{eqnarray} 
with $v^*_{i,q_i}=v_{i,-q_i}$.
The above results for the center-of-mass steady state current and velocity hold as long as the adiabatic approximation (\ref{eq:y:adiab}) is valid.
Consequently,  expanding the ratio $h(x)/g^2(x)$ to leading order in $\epsilon=1/k$ we obtain 

\begin{widetext}
\begin{eqnarray}
U(x)&=& -\int_0^x \left\{ \frac{2 (F_1(y)+F_2(y))}{T_1+T_2}+\epsilon \frac{1}{4  (T_1+T_2)}\partial_y (F_1(y)-F_2(y))^2 \right.\nonumber \\
&&\qquad \qquad\left. +\epsilon\frac{(T_1-T_2)}{  (T_1+T_2)^2}\pq{\frac 1 2 \partial_y(F^2_2(y)-F^2_1(y))+F_1(y) F'_2(y)-F_2(y) F'_1(y)}\right\}\, dy
\nonumber
\\
&=& \frac{2 (V_1(x)+V_2(x))}{T_1+T_2}+\epsilon\frac{\pq{F_1(x)-F_2(x)} \pq{F_1(x) (T_1-3 T_2)+F_2(x) (3 T_1-T_2)}}{4 (T_1+T_2)^2}\nonumber \\
&& -\epsilon \frac{(T_1-T_2)}{(T_1+T_2)^2}\int_0^x F_1(y) F'_2(y)-F_2(y) F'_1(y) \, dy.
\label{eq:U_k}
\end{eqnarray} 
Isolating the periodic part, and introducing $q=2 \pi n/L$, we obtain
\begin{eqnarray}
U_0(x)&=& \frac{2 (V_1(x)+V_2(x))}{T_1+T_2}+\epsilon\frac{\pq{F_1(x)-F_2(x)} \pq{F_1(x) (T_1-3 T_2)+F_2(x) (3 T_1-T_2)}}{4 (T_1+T_2)^2}\nonumber \\
&&-\epsilon\frac{(T_1-T_2)}{(T_1+T_2)^2} \sum_{q\neq -q'} q q' \frac{(q-q')}{(q+q')}v_{1,q} v_{2,q'} \E^{\ii (q+q')x},
\label{U0:eq}
\end{eqnarray}
\end{widetext}
 we can express the effective potential as follows
\begin{equation}
U(x)=U_0(x)-x \bar f,
\label{eq:U_k1}
\end{equation} 
where 
\begin{eqnarray}
\bar f&=&\epsilon \frac{(T_1-T_2)}{(T_1+T_2)^2} \ii \sum_q q^3 (v_{1,q} v_{2,-q}-v_{1,-q} v_{2,q}) \nonumber \\
&=&-2 \epsilon \frac{(T_1-T_2)}{(T_1+T_2)^2} \sum_q q^3\mathrm{Im}(v_{1,q} v^*_{2,q}).
\end{eqnarray} 
In eqs.(\ref{eq:U_k}) and (\ref{U0:eq}) we have omitted the integration constants, as they amount to a constant shift in the effective potential $U_0(x)$.

For potentials $V_1(x)$ and $V_2(x)$ with identical form but shifted $\phi$ with respect to one another, i.e.,
\begin{eqnarray}
V_1(x)&=&\sum_{q}\E^{\ii q x}v_{q},\\
V_2(x)&=&\sum_{q}\E^{\ii q( x+\phi)}v_{q},
\end{eqnarray} 
 we obtain for the effective force 
\begin{equation}
\bar f=2\epsilon\frac{(T_1-T_2) }{ (T_1+T_2)^2}   \sum_{q} |v_{q}|^2 q^3  \sin (q\phi).
\end{equation}

\section{Scaling behavior}
\label{app3}
The final issue is the scaling behaviour of the center of mass velocity and of the heat rates in case the potentials have 
the same period $L$.
The average velocity and the heat rate are given by
\begin{eqnarray}
\bar v&=&\frac 1 2 \average{\dot x_1+\dot x_2}\nonumber \\
&=&-\frac 1 2 \int_0^L \int_0^L \D x_1 \D x_2 \p{V_1'(x_1)+V_2'(x_2)} P(x_1,x_2),\nonumber\\
\average{\dot Q_i}&=& \int_0^L \int_0^L \D x_1 \D x_2 \pq{T_i \partial_i^2 V-\p{\partial_i V}^2} P(x_1,x,_2).\nonumber
\end{eqnarray} 
Introducing the rescaled  coordinates $y_i=x_i/L$
we obtain 
\begin{eqnarray}
\bar v&=&-\frac 1 {2L} \int_0^1 \int_0^1 \D y_1 \D y_2 \p{V_1'(y_1)+V_2'(y_2)} \tilde P(y_1,y_2),\nonumber \\
\average{\dot Q_i}&=&\frac 1 { L^2} \int_0^1\int_0^1 \D y_1 \D y_2 \pq{T_i \partial_{y_i}^2 V-\p{\partial_{y_i} V}^2} \tilde P(y_1,y_2),\nonumber
\end{eqnarray} 
where the conservation of the normalization determines the expression for the rescaled distribution $\tilde P(y_1,y_2)=L^2 P(L y_1,L y_2)$.
If a constant force is applied, e.g., on particle 1, the total potential becomes $V(x_1,x_2)-f_1 x_1$, and by repeating the above analysis, the velocity and heat rate have the same scaling behaviour, provided that the force is rescaled $\tilde f_1=L f_1$.
Thus, if all the other parameters are kept constant, the efficiency (\ref{eta:eq}) as a function of the rescaled force $\tilde f_1$ will be independent of $L$. In particular, its maximum value will not depend on the period $L$.
\end{document}